# Sharing Construction Safety Inspection Experiences and Site-Specific Knowledge through XR-Augmented Visual Assistance

Pengkun Liu[1*], Jinding Xing[1*], Ruoxin Xiong[1] and Pingbo Tang[1**]

*Abstract*— Early identification of on-site hazards is crucial for accident prevention in the construction industry. Currently, the construction industry relies on experienced safety advisors (SAs) to identify site hazards and generate mitigation measures to guide field workers. However, more than half of the site hazards remain unrecognized due to the lack of field experience or site-specific knowledge of some SAs. To address these limitations, this study proposed an Extended Reality (XR)-augmented visual assistance framework, including Virtual Reality (VR) and Augmented Reality (AR), that enables capturing and transferring subconscious inspection strategies between workers or workers/machines for a construction safety inspection. The purpose is to enhance SA's training and real-time situational awareness for identifying on-site hazards while reducing their mental workloads.

## I. INTRODUCTION

The construction site is one of the most hazardous places. In the U.S., among all the fatalities in the private sector, more than 21% of fatalities occurred in the construction industry in 2020 [1]. Besides the fatal injuries, the non-fatal injury rate in the construction industry is still around 29% higher than in the other industries [1]. The cost of these safety incidents exceeds $170 billion each year alone in the U.S [2]. Besides the enormous economic loss, these safety incidents can result in project schedule delays and loss of reputation damage for contractors.

In current practice, safety advisors (SA) perform a routine on-site inspection to identify potential safety hazards and generate mitigation measures accordingly. During on-site visits, the SA firstly analyzes the current safety plans, then walks around the job site to check the current safety situation of the workers, materials, and equipment. After the visit, the SA will report the visit and update the safety plans to mitigate unsafe situations and practices [3-5].

However, previous studies revealed that about 57% of the construction hazards remain unrecognized on the job site [6]. The SA's limited experience or lack of familiarity with certain workspaces or certain parts of domain knowledge are possible contributors. The insufficient number of experienced SAs and the dynamic nature of construction sites further increase SAs' misses of on-site hazards [7].

*Contributes Equally
\*\* Corresponding Author
Pengkun Liu, Jinding Xing, Ruoxin Xiong and Pingbo Tang is with the Department of Civil and Environmental Engineering, Carnegie Mellon University; e-mail: pengkunl, jindingx, ruoxinx, ptang@andrew.cmu.edu.

Extended Reality (XR), which is an emerging technology that melds Augmented Reality (AR) and Virtual Reality (VR), shows the potential to help SA identify on-site hazards effectively. In XR, the SA can navigate or control a digital robot to navigate a virtual construction site (VR) and interact with a virtual interface that contains safety-related information in real construction workspaces (AR). Existing XR-related solutions rely on computer simulations and computer vision techniques to infer the on-site hazards [8]. Current studies provide limited information on less salient on-site hazards, such as hazards associated with manual material handling. The less salient on-site hazards are often interdependences with the construction schedule, physical space, and material properties [9]. In some cases, identifying less salient on-site hazards requires the SA has in-depth safety-related knowledge, rich experiences, and site-specific knowledge. Nevertheless, collecting and reusing experts' safety-related knowledge and site inspection experiences to identify less salient on-site hazards has not been fully explored.

This study presents an XR-augmented visual assistance framework that incorporates the benefits of VR, expert knowledge, and AR to enhance the SA's situation awareness in the safety inspection of the construction site. The proposed framework consists of three major components: (1) VR environment for capturing expert's visual trajectories in safety inspection; (2) inspection strategy discovery through process mining of the captured visual trajectories of a digital drone; (3) AR user interface for visualization inspection strategies.

The contribution of this study to the body of knowledge is three folds: (1) review existing studies relevant to XR and point out the research gaps; (2) propose a framework that captures and shares expert site-specific knowledge and inspection experiences through process mining and XR for assisting SA's safety inspection on construction sites; (3) provide a proof-of-concept model for XR-augmented visual assistance to the SA through a case study of fire safety equipment inspection with a digital drone.

## II. RELATED WORK

This section reviews recent work that uses new technologies to enhance safety inspection on site.

### A. Virtual Reality for Construction

VR is a graphic representation technology that generates a three-dimensional representation of the construction field. Many studies used VR to enhance SAs' safety inspection knowledge by asking the SAs to perform inspection tasks on a virtual construction site [10, 11]. Particularly, the VR model generated from the as-built 3D point cloud provides the SA with a full experience of the real construction site and can offer an effective learning experience for the SA [12]. However, the

VR system does not provide any dynamic highlights to guide SAs in discovering safety-critical objects in given scenes. As a result, the SA needs to interpret the scenarios in the virtual construction site when performing an inspection on a real site, which adds extra cognitive workload to the SA.

### B. Augmented Reality for Construction

AR is a promising approach to mitigate the limitations of VR. Some other studies used AR as a 3D viewer that overlays virtual objects and information on the physical site views to help SA identify and recognize hazards [13, 14]. The virtual information generated from building information modeling (BIM) provides rich geometry information about the construction project [15]. Such methods help the SA maintain better situational awareness of the construction site. Yet, the virtual information generated from computer simulations of a construction environment provides a limited reflection of real-world environments, such as time, physical space, and material properties [9]. Often the virtual information helps the SA infer geometry-related hazards, such as falls, struct by an object [16]. It remains challenging for naive SA to identify less salient hazards, such as the spreading of chemical hazards and fires.

### C. Extended Reality for Construction

Some studies used a combination of XR and machine learning techniques to deliver and visualize hazard information to SAs [16]. These studies use an AR interface to visualize information relevant to site hazards detection. The virtual information is generated through computer vision techniques that detect hazardous scenarios. Navigation planning algorithms are also used to generate a safe and efficient path for SAs [17]. The combined solution for improving safety inspection on-site has been provided to be very effective. However, this solution overlooked experts' experiences and site-specific knowledge, which is critical to identifying less salient hazards in dynamic and cluttered environments.

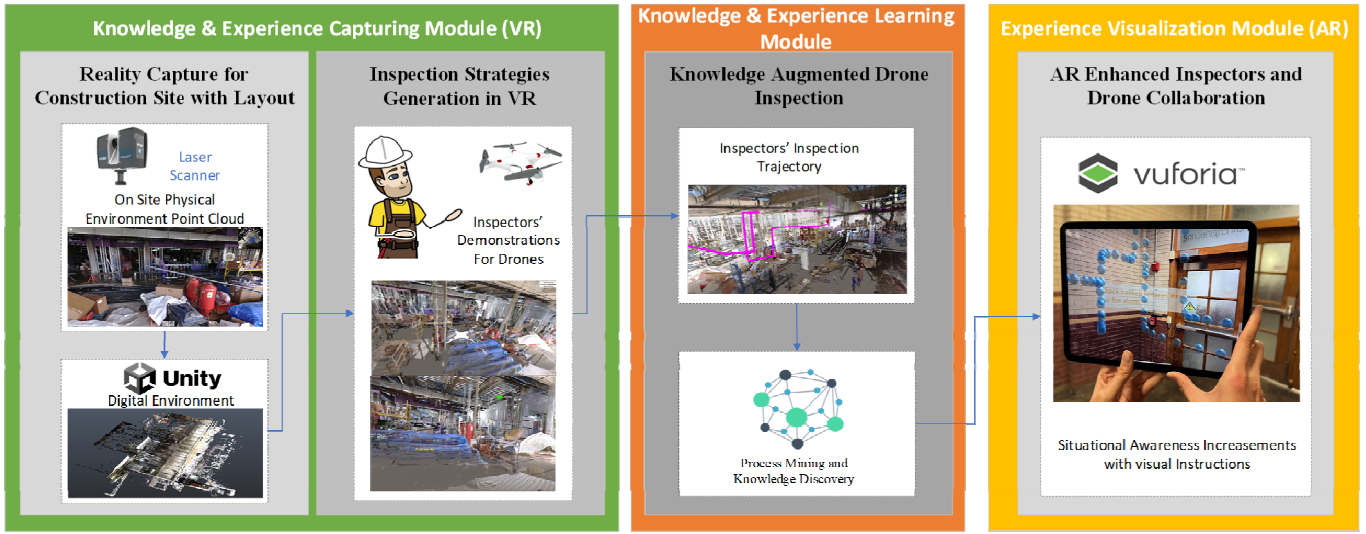

Figure 1. The proposed XR-augmented visual assistance framework for assisting construction safety inspection

## III. PROPOSED SOLUTION

This study proposed an XR-augmented visual assistance system aiming to learn and transfer implicit inspection strategies associated with subconscious experiences and site-specific knowledge between SAs or SAs/machines for a construction safety inspection to enhance SA's situational awareness. The proposed system comprises three modules, as shown in Figure 1: (1) inspection visual trajectory capture module (VR module): where a SA's behaviors and inspection strategies are recorded through controlling a digital drone during the interaction with the reconstructed digital environment in the VR environment; (2) inspection knowledge & experience learning module (learning module), where process mining algorithms will be applied to discover knowledge and inspection strategies captured from the VR module; (3) and a visualization interface (AR interface), where AR-enhanced SAs and digital drone collaborate to detect real-hazards or hazardous scenarios with unsafe combinations of objects' states. Figure 1 shows the overall framework of the proposed system.

The VR module aims to capture SA's observation trajectories in identifying on-site construction hazards. The inspection knowledge and experience capturing module includes a virtual construction site generated from a 3D scan of the construction site and a virtual drone. The SA can control a digital drone from the first perspective to observe and interact with the virtual construction site in the VR module, as shown in Figure 2. Meanwhile, the virtual drone will record the SA's observation trajectories in the VR environment.

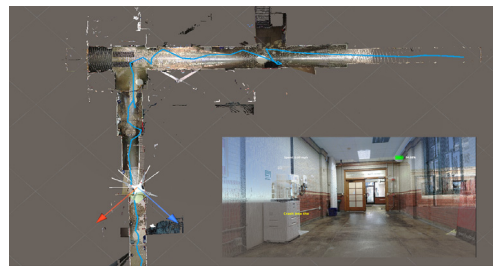

Figure 2. Safety inspectors use a digital drone to explore and search hazards on the virtual construction site

The knowledge and experience learning module is tasked with collecting multiple SAs' inspection strategies, especially the object detection sequences and related contextual information. As shown in Figure 3, the digital drone equipped with laser for fire safety equipment detection, such as the emergency exit lights, fire alarm panels and so on. Furthermore, the point cloud segmentation for scene understanding and semantic mapping with building information modeling is also significant for learning such inspection strategies. Then the knowledge of inspection of SA's observations and related decisions cloud be stored as a knowledge graph [18-20].

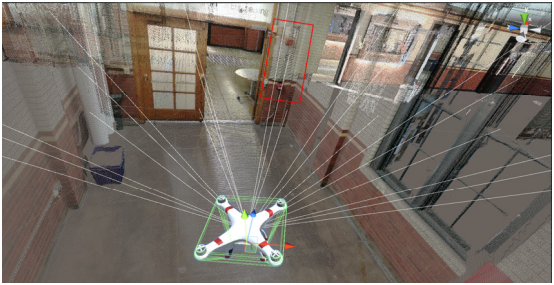

Figure 3. *Digital drone equipped with laser for object detection (emergency exit lights and fire alarm panels)*

The AR interface is used to visualize and replay the learned inspection strategies and knowledge. This paper developed an iOS app by Vuforia to facilitate the SA's safety inspection process. In the app, augmented information (expert's observation trajectories and relevant safety information) overlay in the real construction site to direct the SAs for identifying on-site hazards and increasing their situational awareness.

## IV. CASE STUDY

This study used a campus building to test the proposed system. The participants are asked to perform fire safety inspections inside the building. During the inspection process, SAs should check the status of the fire extinguishers, fire alarms, exit signs, and fire rescue equipment as required by the current building safety regulations. Missed potential safety hazards can lead to devastating consequences in the fire situation. The goal of the inspection task is to ensure that the fire equipment is in functional conditions.

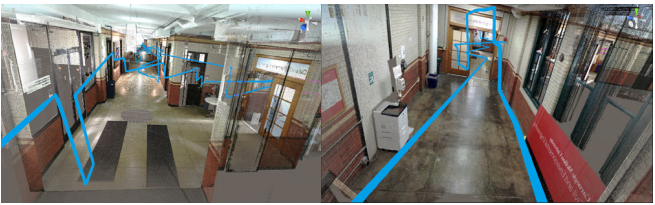

Figure 4. *Visualization of SA's inspection histories*

This study used iPad pro 11 to capture the 3D layout of the campus building and visualize the generated fire inspection information. The first phase is to create a digital VR model from the physical world in Unity 3D. Given that, the laser scanner is used to capture the high-quality point cloud of a campus building. Once the point cloud of the physical world is collected, it will be imported into Unity 3D to rebuild the digital environment. Then in the digital environment, the SA can navigate in the environment and carry out fire equipment inspection without any guidance. Then the SA's inspection strategies would be recorded through the first-person perspective with timestamps, as shown in Figure 2. The blue lines demonstrate the SA's inspection strategies and attention division. Especially, inspectors would spend more time checking the fire equipment and switching the camera angle to look around the surrounding environments. Therefore, the blue lines represent the inspection strategies which would be denser and clustered near the fire equipment. Once the strategies are collected, the SA's inspection visual trajectory will be used to guide the digital drone to navigate and search the objects in the digital environment, as shown in Figure 3. In addition, to make the digital drone capable of sensing, the digital drone is equipped with multiple laser sensors to detect the objects in the digital environment, as shown in Figure 4. Therefore, the collected inspection trajectories are mapped with semantic information to understand the mechanics of SAs' observations and decisions.

Finally, with the help of AR, in the physical world, the on-site SA equipped with an iPad could replay the inspection strategies and check the inspection instructions, as shown in Figure 5. The optimal trajectories and instructions representing expert knowledge could be delivered to naive SAs. For example, in the fire extinguisher inspection task, the AR module maps the visual inspection trajectory in the reconstructed VR environment (the trajectory constitutes of blue spheres) to the real world. Such mapping can also show detailed inspection instructions, such as checking the test and maintenance dates. Furthermore, the density of the blue spheres represents the relative attention time, which means the SA spends more time at this place to check the equipment, such as the SA would check the battery conditions of fire alarms and exit signs/lights.

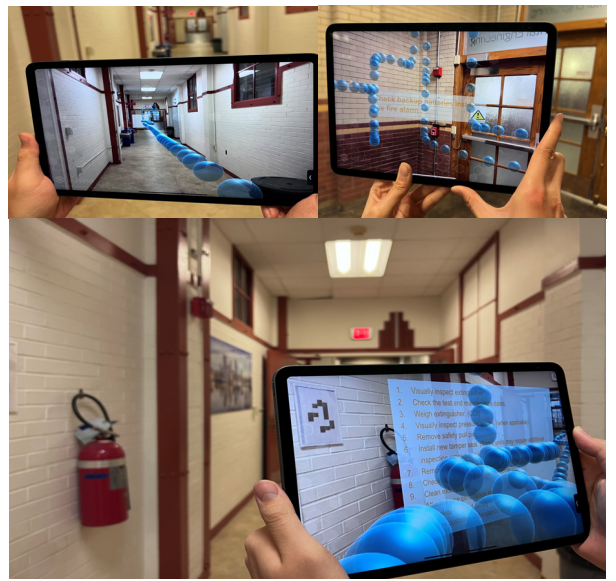

Figure 5. An inspector navigates to target equipment with the XR-augmented visual assistance platform

Therefore, the authors demonstrate the potential of the proposed XR-augmented system for the fire safety equipment inspection tasks in a physical-digital environment where a digital drone collects and learn from the SA's inspection strategies and replays the searching strategies for SAs in the physical environment with detailed instructions.

## V. CONCLUSION AND FUTURE WORK

The authors demonstrated the potential for increasing the inspectors' situational awareness in a fire equipment safety inspection task with XR. In the case study, the SA perform inspection tasks by controlling a digital drone to navigate in a VR construction site. The digital drone could learn from the SA's inspection strategies and localizes potential safety hazards to assist safety inspectors to identify field hazards. Furthermore, the AR module could replay the inspection strategies in the physical world to enhance the SAs' situational awareness.

This paper has not fully considered the process mining and knowledge discovery from the multiple inspectors' observation trajectories. For future work, this paper will pattern mining of the SAs' observation trajectories collected in the VR world and explore the relationship between the mined pattern and the BIM model to summarize the knowledge of SAs.

## ACKNOWLEDGMENT

This material is based on work supported by the Nuclear Energy University Program (NEUP) of the U.S. Department of Energy (DOE) under Award No. DE-NE0008864, and NASA University Leadership Initiative program (Contract No. NNX17AJ86A, Project Officer: Dr. Anupa Bajwa, Principal Investigator: Dr. Yongming Liu).